\documentclass[%
longbibliography,
 prfluids,
 amsmath,amssymb,
 aps
]{revtex4-2}

\usepackage{graphicx}
\usepackage{dcolumn}
\usepackage{bm}
\usepackage{hyperref}

\usepackage{xcolor}

\newcommand{\ecrit}{e_\mathrm{crit}}

\newcommand{\od}[1]{\mathcal{O}\left(\epsilon^{#1}\right)}
\newcommand{\q}[2]
{
   \frac{\partial #1}{\partial #2}
}

\newcommand{\ud}{\,\mathrm{d}}
\newcommand{\tend}{{t_\mathrm{f}}}
\newcommand{\hreg}{
H_\text{reg}
}
\newcommand{\gammaReg} {
\gamma_\text{reg}
}
\newcommand{\hterm}{
H_\text{term}
}
\newcommand{\gammaterm}{
\gamma_\text{term}
}
\newcommand{\Eb}{
{}
}

\begin{document}

\title{Electrostatic control of the Navier--Stokes equations for thin films} 

\author{Alexander W. Wray}
 \email{alexander.wray@strath.ac.uk}
\affiliation{%
Department of Mathematics and Statistics, University of Strathclyde, Livingstone Tower,
26 Richmond Street, Glasgow G1 1XH, UK
}%

\author{Radu Cimpeanu}
 \email{radu.cimpeanu@warwick.ac.uk}
\author{Susana N. Gomes}
 \email{susana.gomes@warwick.ac.uk}
\affiliation{
Mathematics Institute, University of Warwick, Coventry
CV4 7AL, UK
}

\date{\today}

\begin{abstract}
A robust control scheme is derived and tested for the Navier--Stokes equations for two-dimensional multiphase flow of a thin film underneath an inclined solid surface. Control is exerted via the use of an electrode parallel to the substrate, which induces an electric field in the gas phase, and a resultant Maxwell stress at the liquid-gas interface. The imposed potential at the second electrode is derived using a Model Predictive Control loop, together with Optimal Control of a high-fidelity reduced-dimensional model. In this implementation the interfacial shape of the fluid is successfully controlled, however the algorithm is sufficiently general to control any other quantity of interest.
\end{abstract}

\maketitle

\section{Introduction}

Thin films find applications in numerous situations, from lubrication systems to printing \citep{oron1997long,craster2009dynamics,kalliadasis2011falling}. 
However, while many applications involve control (e.g. of interfacial shape), achieving this systematically has remained relatively unexplored until recently, with most efforts focusing on the effect of varying different external mechanisms (e.g. electric fields or thermal effects), rather than doing so in a theoretically grounded and robust way - an omission we aim to resolve. 

A common approach to understanding thin films is to use a {\em reduced-dimensional} model (RDM).
Classical RDMs exploit lengthscale separations in problems (e.g. liquid film aspect ratios) to derive comparatively simpler evolution equations for the film height $h$ \citep{craster2009dynamics,kalliadasis2011falling}, typically reducing the dimensionality of the system by one. 
More modern models, such as the Weighted Residual Integral Boundary Layer (WRIBL) model \citep{ruyer2000improved}, have introduced an additional degree of freedom - the flux $q$ - ultimately yielding quantitative agreement with Direct Numerical Simulations (DNS) and experiments \citep{scheid2006wave,chakraborty2014extreme}. 
Such RDMs in 1+1 dimensions (i.e. one spatial and one temporal dimension) on fixed domains are significantly faster to solve than the corresponding 2+1 dimensional (two spatial dimensions and one temporal dimension) DNS on free-boundary domains. 
Here we confine our attention to two spatial dimensions 
to ensure tractability: the numerous computations involved (see Section~\ref{sec:control}) would require specialist computational resources in 3+1D (or 2+1D for the corresponding RDMs).

Many actuation mechanisms could suffice for controlling films, including blowing and suction \citep{thompson2016falling} or thermocapillarity \citep{thompson2019robust}.
Each physically-driven actuation requires additional components for the model, generally based on a WRIBL-like projection method \citep{thompson2016falling,wray2017accurate}.
While control algorithms are essentially actuation-independent, here electric fields are used, due to the comparative ease of machining finely-controllable actuators \cite{thompson2019robust}.
%
Although a common model for electrostatic systems is the {\em leaky-dielectric} model \citep{saville1997electrohydrodynamics,papageorgiou2019film}, we use a simplification thereof in which the liquid is perfectly conducting.
%
%

Most theoretical control works focus on linear PDEs or scenarios in which the nonlinearity is prescriptive (e.g. Lipschitz continuous) due to their analytic tractability \citep{troltzsch2010optimal}. As a result, the only models of fluid flows that have received analytical control treatment have been based on linear \citep{cerpa2017control} or weakly nonlinear \cite{gomes2017stabilizing,armaou2000feedback,armaou2000wave,al2018linearized,tomlin2019point} equations. More realistic, but more highly nonlinear, models have been examined using numerical machine learning \citep{belus2019exploiting} or optimal control (OC) techniques \citep{sellier2016inverse,boujo2019pancake}, but, at least for DNS, these have only been for single-phase flows \cite{sipp2016linear}. 
We are interested in multiphase flows, where applying OC to DNS would be prohibitively costly. Moreover, applying controls derived from RDMs fails due to accumulated error between the RDM and DNS over time \citep{cimpeanu2021active}.
High performance computing tools have accelerated our understanding of thin-film flows and their control:
from extending RDM limitations into inertial regimes \citep{denner2018solitary,wray2020reduced}, to explaining phenomena such as dripping \cite{rohlfs2017hydrodynamic, kofman2018prediction} and extending into multi-physics contexts such as the study of electrostatic instabilities \cite{anderson2017electric, cimpeanu2014control, tomlin2019point}, the ability to inspect flow quantities in spatiotemporal detail has proven invaluable.
Some progress has already been made using {\em feedback control} by applying the results of linear stability analyses to more complex problems \citep{thompson2016stabilising,gomes2017stabilizing}. \citet{cimpeanu2021active} extended this by applying the {\em control rules} derived from linear theory (i.e. imposing equivalent proportional controls) to successfully drive the system towards a desired state. This has motivated the exploration of the {\em Model Predictive Control} (MPC) methodology herein.
MPC consists of finding an updated control for a system at discrete time points during its evolution \cite{grune2017nonlinear}. Typically this is achieved by performing a control calculation over a shorter time horizon. We propose to use this methodology, but to derive the control from a RDM initiated with suitable measurements from the DNS. We will show that this methodology is robust, avoiding issues with  both model errors and numerical errors.


The methodology described herein should form the first step towards practically-realisable online control of real-world systems. 
Using DNS as a high-fidelity digital analogue for experiments, these techniques are demonstrated to afford sufficient accuracy and specificity to exert fine control over a fluid, including, for example, its precise interfacial shape. 
While future developments are yet needed, we anticipate that the methodology presented here will underpin practical implementations of techniques as diverse as 3D printing and minimising carbon footprints.

\section{Modelling}
\subsection{Governing equations}

\begin{figure}
\centering
\includegraphics[width=0.75\textwidth]{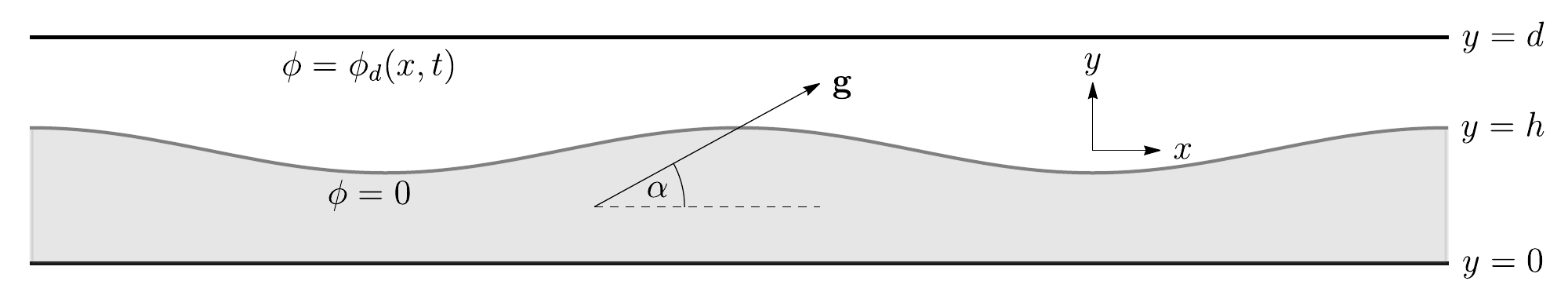}
\caption{Geometry of the hanging liquid film described by interfacial height $h(x,t)$, with the multi-fluid system confined between two electrodes placed at $y=0$ and $y=d$, at inclination angle $\alpha$. All quantities are dimensionless.}
\label{fig:schematic}
\end{figure}

We consider a perfectly-conducting Newtonian fluid of constant density $\rho$ and viscosity $\mu$ hanging from a planar electrode inclined at an angle $\alpha$, held at zero potential without loss of generality, as shown in Figure \ref{fig:schematic}. We work in Cartesian coordinates $\mathbf{x}=(x,y)$. A second parallel electrode at $y=d$ has potential $\phi_d(x,t)$. An electric potential $\phi$ is induced in the gas phase, which has permittivity $\epsilon_G$, and couples into the hydrodynamic problem via the addition of the Maxwell stress tensor $\mathcal{M}$ to the total stress tensor $T$. The velocity and pressure in the fluid are given by $\mathbf{u}=(u,v)$ and $p$, respectively. The liquid-gas interface has constant surface tension coefficient $\sigma$. The system is nondimensionalised as
\begin{equation}
    \mathbf{u}=U\mathbf{\hat{u}}, \quad \mathbf{x}=H\mathbf{\hat{x}}, \quad p=\frac{U\mu}{H}\hat{p}, \quad t=\frac{H}{U}\hat{t}, \quad \phi=\phi_b\hat{\phi}, \quad U=\frac{\rho g H^2}{\mu}, \quad \phi_b=\sqrt{\frac{HU\mu}{\epsilon_G}},
\end{equation}
where $H$ is the undisturbed film thickness, $U$ is a characteristic velocity, and $\phi_b$ is a characteristic potential. 
The hydrodynamic part of the problem is governed by the standard Navier--Stokes and continuity equations, subject to no-slip and impermeability conditions at the wall $y = 0$. 
At the interface $y = h(x,t)$, the usual kinematic and tangential stress conditions are unchanged, while the normal stress condition becomes
\begin{equation}
    p-\Gamma\kappa=\frac{2}{1+h_x^2}\left( v_y+h_x^2u_x-h_x\left(v_x+u_y\right) \right)-\Eb E^N, \quad E^N=\frac{1}{2}\left(\phi_x^2-\phi_y^2\right)(h_x^2-1)-2h_x\phi_x\phi_y,\label{eq:ENEq}
\end{equation}
where $\Gamma=\frac{\sigma}{\mu U}$ is the inverse capillary number measuring the relative strength of surface tension, $Re=\frac{\rho U H}{\mu}$ is the usual Reynolds number, and $\kappa$ is the interfacial curvature. 
The potential in the gas, $\phi$, satisfies Laplace's equation \citep{pillai2018nonlinear,craster2005electrically} subject to suitable equipotentials at the interface and the second electrode,
\begin{equation}
    \nabla^2\phi=0 \quad \text{s.t.}\quad \phi|_{y=h}=0, \quad \phi|_{y=d}=\phi_d. \label{eq:fullLapBC}
\end{equation}

While an experiment would be performed under open-flow conditions, 
here we chose to use periodic boundary conditions in the stream-wise ($x-$) direction. 
This can be justified by noting that for sufficiently long domains (here $L=30$), the interface is essentially fully developed into a saturated travelling wave state before it travels one period, and therefore we are not introducing unrealistic behaviour by considering these boundary conditions. For controls derived using linear stability analysis, the boundary conditions affect the resulting controls (and one's ability to justify them theoretically); however, in the OC case presented here, an open flow situation would only alter the methodology via the incorporation of open flow boundary conditions in the adjoint equations below. 

\subsection{Reduced-dimensional model}\label{sec:RDM}
The control scheme derived in Section~\ref{sec:control} will make extensive use of a RDM for speed of computation. Therefore, we aim to derive a high-accuracy model which can be computed quickly. We derive a long-wave asymptotic model using the standard substitutions $\partial_x\mapsto\epsilon\partial_x,\ \partial_t\mapsto\epsilon\partial_t,\ v\mapsto\epsilon v$, where $\epsilon$ indicates the relative sizes of the respective terms. 
%
We use a simplified second-order WRIBL model following \citet{wray2017accurate} for the hydrodynamic component of the model to yield
\begin{align}
    h_t+q_x&=0,\label{eq:finalKinEq}\\
    \epsilon Re\,q_t&=\epsilon Re\left(\frac{9}{7}\frac{q^2}{h^2}h_x-\frac{17}{7}\frac{q}{h}q_x\right)+\frac{5}{6}h\sin\alpha-\frac{5}{2}\frac{q}{h^2}+\epsilon\frac{5}{6}\Gamma h h_{xxx}+\epsilon\frac{5}{6}\cos\alpha h h_x+\epsilon\frac{5}{6}\Eb  h E^N_x\label{eq:finalEvEq}\\
&+\epsilon^24\frac{q}{h^2}h_x^2+\epsilon^2\frac{9}{2}q_{xx}-\epsilon^2\frac{9}{2}\frac{1}{h}q_xh_x-\epsilon^26\frac{q}{h}h_{xx}. \nonumber
\end{align}

At leading order, the electric field satisfies
\begin{equation}
\phi_{yy}=0 \quad \text{s.t.} \quad \phi|_{y=0}=0, \quad \phi|_{y=d}=\phi_d, \quad \text{with solution} \quad \phi=\phi_d\left( \frac{y-h}{d-h} \right). \label{eq:leadingOrder}
\end{equation}
To proceed to higher order, we project onto
\begin{equation}
\phi=f_0(x,t)+f_1(x,t)(y-h)+\epsilon^2\left[f_2(x,t)(y-h)^2+f_3(x,t)(y-h)^3\right]+\od{4}.
\end{equation}
Imposing Laplace's equation \eqref{eq:fullLapBC} up to second order and the equipotential at the interface $y = h$ \eqref{eq:fullLapBC}, yields
\begin{equation}
\phi=f_1(y-h)\left(1+\epsilon^2h_x^2\right)-\epsilon^2\frac{1}{6}\left[ f_1(y-h)^3 \right]_{xx}+\od{4}.\label{eq:f1Inner}
\end{equation}
The substitution $f=f_1\left(1+\epsilon^2h_x^2\right)$ simplifies this to
\begin{equation}
\phi=(y-h)f-\epsilon^2\frac{1}{6}\left[ (y-h)^3f \right]_{xx}+\od{4},\label{eq:fInner}
\end{equation}
where $f$ is determined by imposing the equipotential at the second electrode $y = d$ \eqref{eq:fullLapBC},
\begin{equation}
(d-h)f-\epsilon^2\frac{1}{6}\left[ (d-h)^3f \right]_{xx}=\phi_d.\label{eq:fInnerCons}
\end{equation}
Substituting \eqref{eq:fInner} into \eqref{eq:ENEq} yields the remarkably simple form $E^N=f^2/2+\od{4}$.  
This model avoids the spurious locality of lubrication-type models \citep{rohlfs2021effect}, better preserving the elliptic nature of the underlying Laplace equation. 
It is found to give excellent agreement with DNS even deep into the shortwave regime - an in-depth discussion of which we defer to a companion paper.

\section{Control framework and results}\label{sec:control}

\subsection{Optimal control of the RDM}\label{sec:ocFramework}

\begin{figure}
\centering
\begin{tabular}{ccc}
\includegraphics[width=0.48\textwidth]{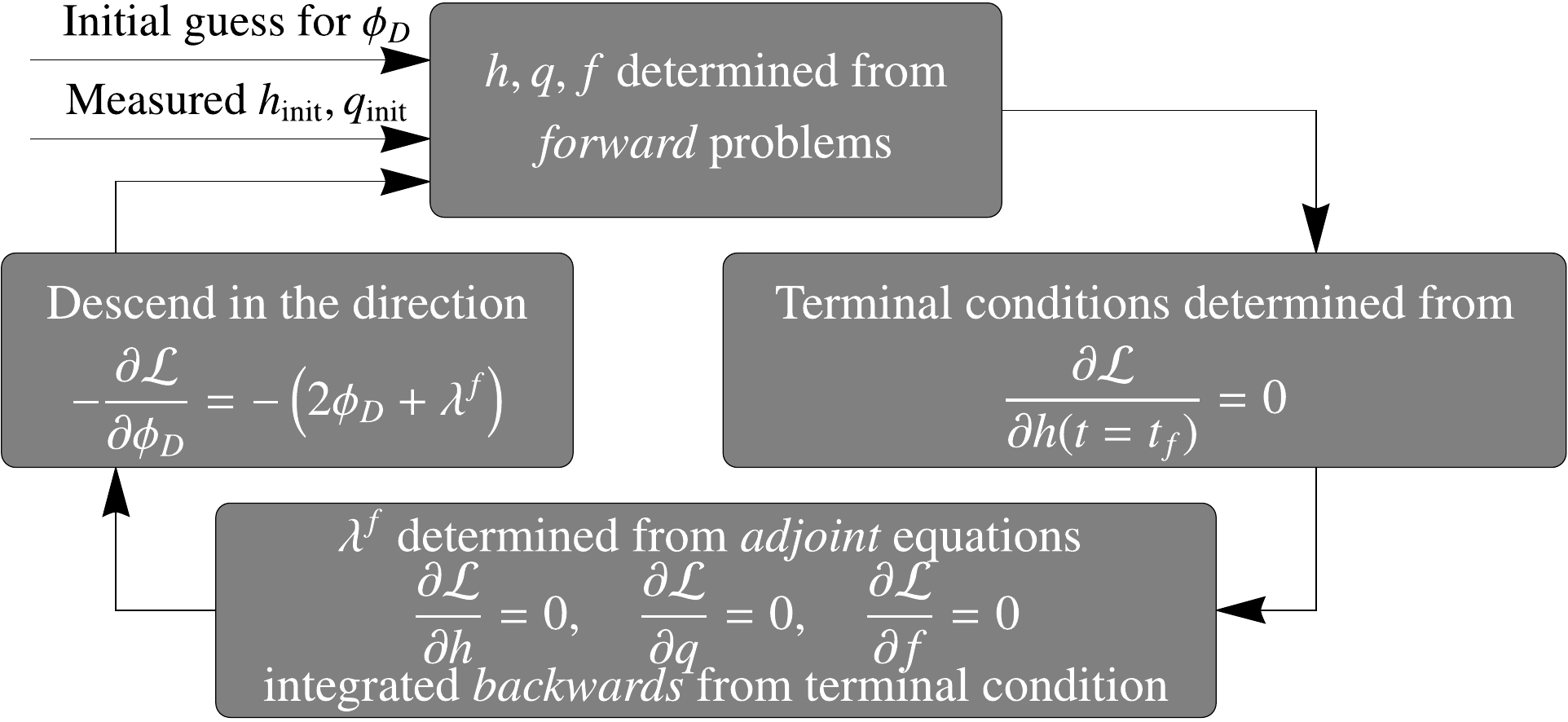}
 & \, &
 \includegraphics[width=0.4\textwidth]{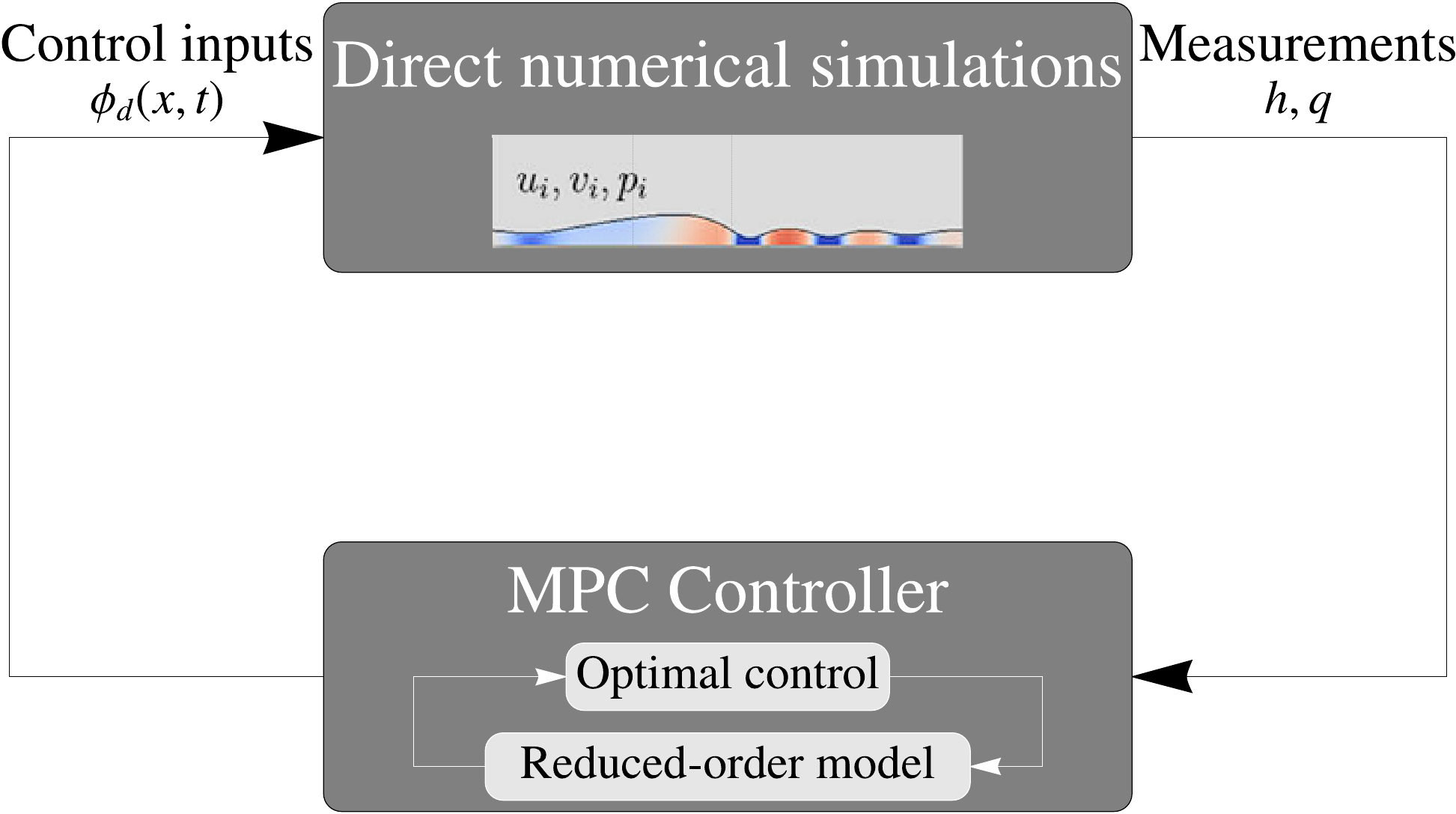}
 \\
(a) && (b) 
\end{tabular}
\caption{
(a) Iterative loop used for OC (forms MPC controller in (b) for control of DNS).
(b) Loop used for MPC.
}  
\label{fig:mpcOC}
\end{figure} 

As OC is becoming more common in fluid-dynamical contexts \cite{sellier2016inverse,boujo2019pancake} we only provide an overview.
We aim to control the interface towards some target state, either at the final time $\tend$, or across the duration of the flow. It is often desirable to achieve this at some minimum cost -- either in terms of energy used, or perhaps in terms of gradients of the control (e.g. to aid manufacturing). Here, we minimise the integral square potential used at the second electrode, introducing the {\em cost functional} $\mathcal{J}$,
\begin{align}
    \mathcal{J}=\gammaReg & \int_{t=0}^\tend \int_{x=0}^L (h(x,t)-\hreg(x,t))^2\, \ud x \ud t+\gammaterm \int_{x=0}^L \left(h(x,\tend)-\hterm(x)\right)^2\,\ud x +\gamma  \int_{t=0}^\tend \int_{x=0}^L \phi_d^2\, \mathrm{d}x\,\mathrm{d}t. \label{eq:costFunc} 
\end{align}
These penalise respectively for:
\begin{enumerate}
    \item Deviation of $h$ from $\hreg$, uniformly weighted across the whole time interval (regulation control).
    \item Deviation of $h$ from $\hterm$ at the final time $\tend$ (terminal control).
    \item Integral square potential used at the second electrode.
\end{enumerate}

Typically we take $(\gammaReg,\gammaterm)=(1,0)$ or $(0,1)$ to impose regulation or terminal control, respectively. 
$\gamma$ controls the relative importance of the control cost compared to the cost of deviating from the desired state. For our present purpose we chose $\gamma=10^{-8}$, which could be considered small when compared to the existing literature \citep{boujo2019pancake}. We chose this value to test the feasibility of our methodology (if an ``infinite" control did not work, the methodology would never be feasible), but retained it after observing that the resulting controls are always realisable (see discussion below). 
The effects of varying this constant would be an interesting path for future study. 
The cost functional \eqref{eq:costFunc} is minimised subject to the constraints that $h$, $q$ and $f$ satisfy the governing equations \eqref{eq:finalKinEq}, \eqref{eq:finalEvEq}, \eqref{eq:fInnerCons}. This is converted into an unconstrained optimisation problem by introducing the Lagrangian
\begin{align}
    \mathcal{L}=\mathcal{J}+\int_{t=0}^T \int_{x=0}^L & -\lambda^h\left\{ h_t+q_x \right\}-\lambda^f\left\{ (d-h)f-\frac{1}{6}\left[\left(d-h\right)^3f\right]_{xx}-\phi_d \right\}\mathrm{d}x\,\mathrm{d}t\\
+\int_{t=0}^T \int_{x=0}^L &-\lambda^q\left\{ Re\left(-q_t+\frac{9}{7}\frac{q^2}{h^2}h_x-\frac{17}{7}\frac{q}{h}q_x\right)+\frac{5}{6}h\sin\alpha-\frac{5}{2}\frac{q}{h^2}+\frac{5}{6}\cos\alpha h h_x\right.\\
&\qquad\qquad\left.+\frac{5}{6}\Gamma h h_{xxx}+\frac{5}{6}\Eb  h f f_x+4\frac{q}{h^2}h_x^2+\frac{9}{2}q_{xx}-\frac{9}{2}\frac{1}{h}q_xh_x-6\frac{q}{h}h_{xx} \right\}\mathrm{d}x\,\mathrm{d}t,
\end{align}
where $\lambda^h(x,t)$, $\lambda^q(x,t)$ and $\lambda^f(x,t)$ are Lagrange multipliers. 
Minimising $\mathcal{L}$ is a standard unconstrained optimisation problem which is solved numerically using conjugate gradients. 
As shown in Figure \ref{fig:mpcOC} (a), at any iteration, the direction in which $\mathcal{L}$ decreases most steeply with respect to the control is
\begin{equation}
    -\q{\mathcal{L}}{\phi_d}=-\left(2\phi_d+\lambda^f\right), \label{eq:descentDirection}
\end{equation}
where functional derivatives are interpreted as Fr\'echet derivatives. 
Computing \eqref{eq:descentDirection} requires $\lambda^f$, which is determined by solving the {\em adjoint} equations obtained from
\begin{equation}
    \q{\mathcal{L}}{h}=0, \quad \q{\mathcal{L}}{q}=0, \quad \q{\mathcal{L}}{f}=0,\label{eq:adjoinEqs}
\end{equation}
leading to time-dependent PDEs for $\lambda^h$ and $\lambda^q$, and a boundary-value equation for $\lambda^f$, presented in Appendix \ref{sec:appOC}. 
These equations are integrated backwards in time from $t=\tend$ to $t=0$ subject to the terminal condition  $\lambda^h(x,\tend)=2\gammaterm(h(x,\tend)-\hterm)$, derived from the constraint $\partial \mathcal{L}/\partial h(x,\tend)=0$. 
Equations \eqref{eq:adjoinEqs} require $h$, $q$ and $f$, which can be determined by standard forwards integration in time.

A Polak-Ribi\`ere \citep{polak1969note} conjugate gradient method 
was used to determine the descent direction, while a golden section method \citep{kiefer1953sequential} was used for the line search. %
The convergence criterion was that $\mathcal{L}$ had not varied by more than $10^{-6}$ across $10$ iterations (see Appendix \ref{sec:appOC} for details).

To prevent the electric fields violating the dielectric limit, the constraint $\phi_d\leq k{\phi_d}_\text{max}$ was imposed, where $k<1$ is a constant, and ${\phi_d}_\text{max}$ is the dimensionless constant potential that would induce an electric field at the dielectric limit. Such saturation constraints have been previously included to account for realistic actuator behaviour \cite{fabbiane2014adaptive}.
In practice, due to the long-wave instabilities, it was sufficient to take $k=0.75$. 
This was imposed by taking suitable projections of both controls and descent directions to prevent constraint violation. 

Not all target states are attainable: regulation control towards sharp, stationary states proves especially problematic due to the convective nature of the system. 
However, producing the correct shape is often more important than its exact location (cf. manufacturing applications). 
To that end, a mechanism was added to allow for translation-agnostic target states. 
After each iteration, the target state for each time step was translated to minimise root-mean-square error between the target state and the latest iteration of the interface. 
This yields target states with the correct shape, but advected with the mean flow, giving much wider ranges of accessible target states.

\begin{figure}[!hb]
\centering
\begin{tabular}{ccc}
\includegraphics[width=0.28\textwidth]{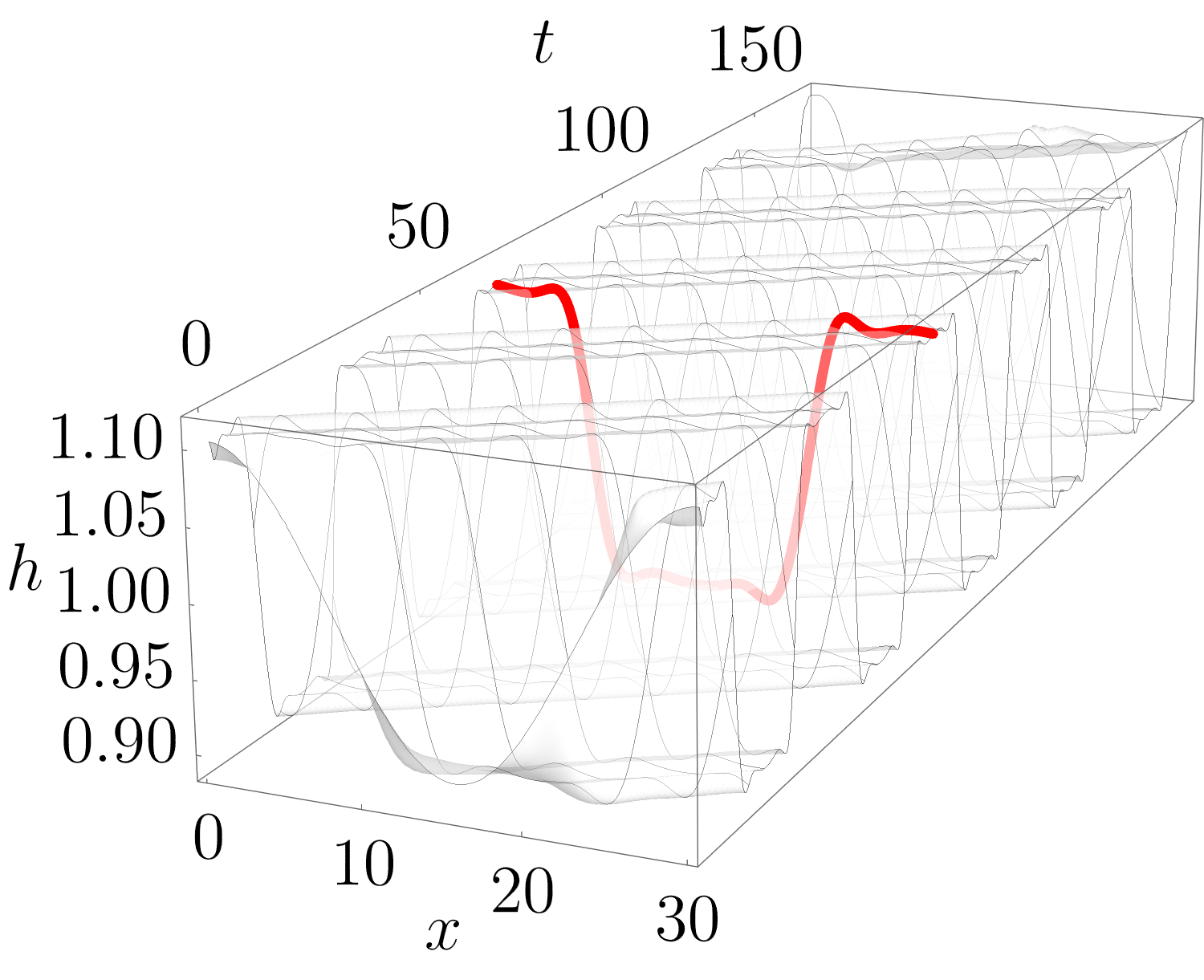} &
\includegraphics[width=0.34\textwidth]{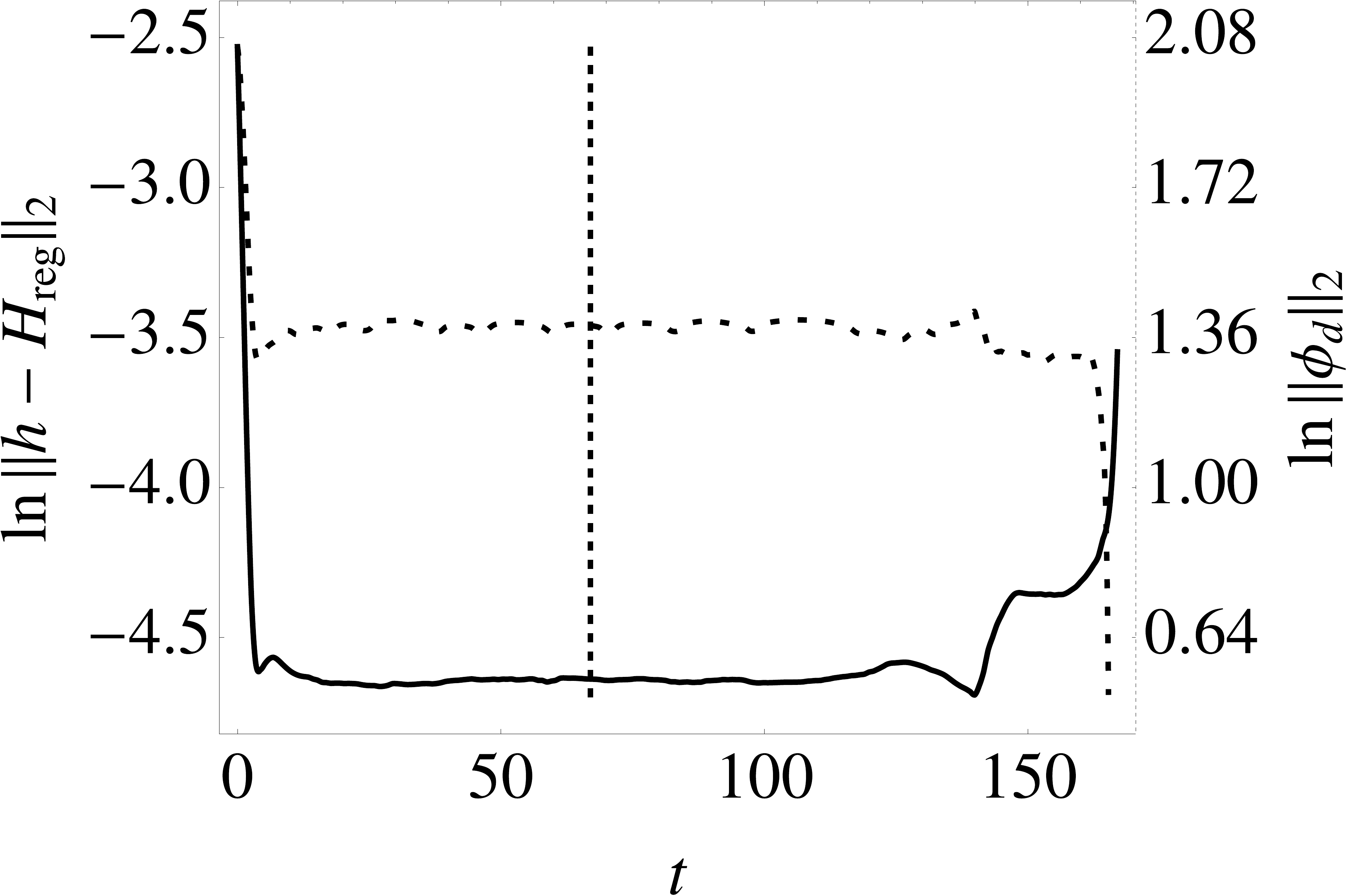} &
\includegraphics[width=0.28\textwidth]{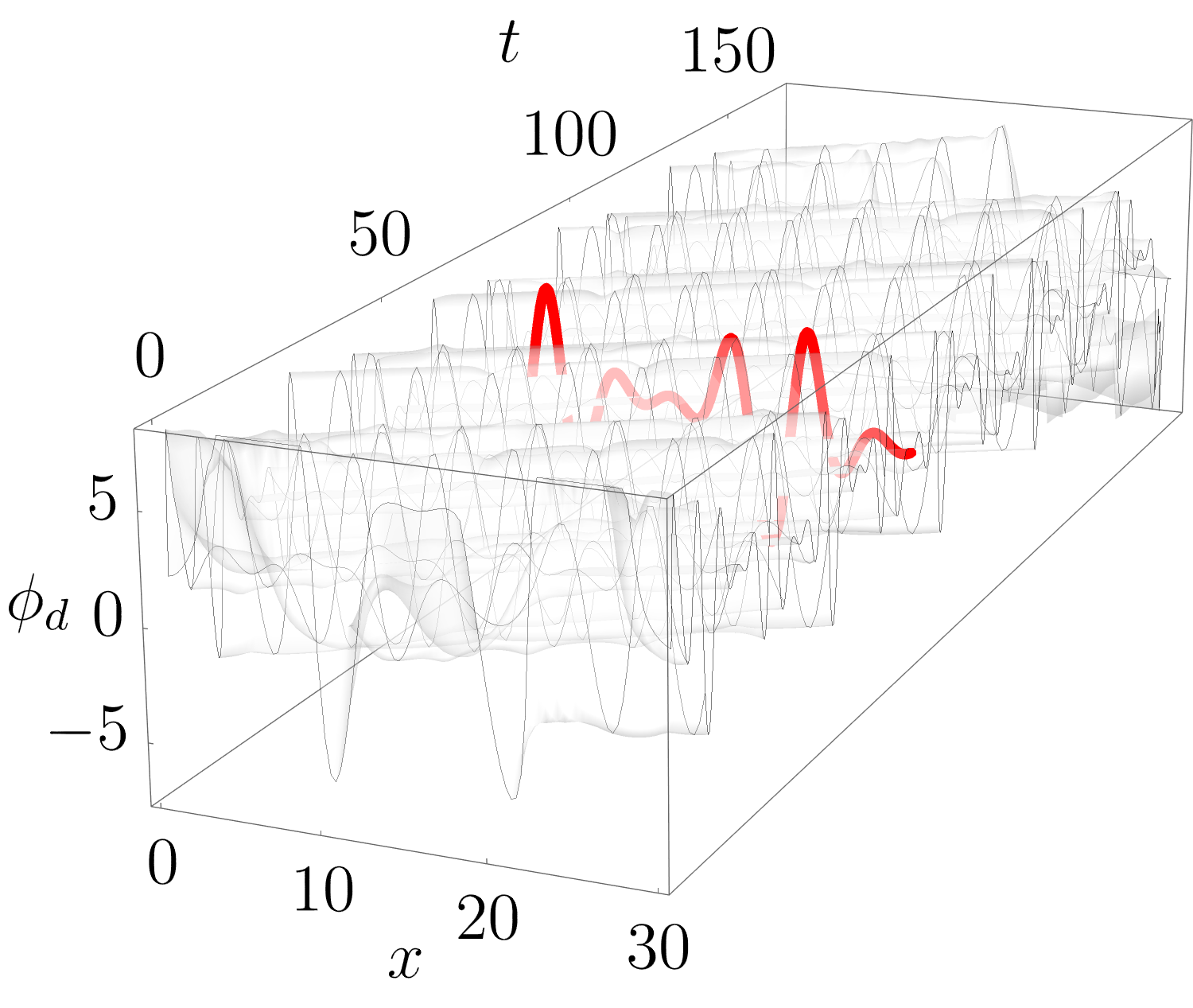}\\
(a) & (b) & (c) \\
\end{tabular}
\caption{Regulation control of the RDM towards \eqref{eq:tophat}. (a) Interface height $h$ over time (highlighted: interface at $t=0.4\tend$); (b) $\ln ||H-H_\mathrm{reg}||_2$ (solid line; left axis) and $\ln||\phi_d||_2$ (dashed line; right axis); vertical dotted line: $t=0.4\tend$; (c) Voltage potential $\phi_d$ over time. }
\label{fig:LOMControl}
\end{figure} 

As an example, in Figure \ref{fig:LOMControl} we examine translation-agnostic regulation OC towards
\begin{equation}
    H_\mathrm{reg}=0.9+0.1(\tanh(x-L/4)-\tanh(x-3L/4)).\label{eq:tophat}
\end{equation}
We use the parameter values
\begin{equation}
\rho=1.1\times10^3\ \text{kg m}^{-3}, \quad \mu=0.1\ \text{kg (m s)}^{-1}, \quad \alpha=3\pi/8, \quad H=1\times10^{-3}\text{ m}, \quad d=5H, \quad \gamma=0.05\text{kg s}^{-2}.   
\end{equation}
The fluid interface is successfully controlled towards the target as reflected in the low error in (b), although (a) shows that this target translates around the (periodic) domain at an approximately constant rate. 
As commonly seen, the error increases in the last stages of the computation. This is because, in these latter stages, there is less value in working to stay close to the target state than the cost saved by switching off the controls, as deviations only matter for a short time; this also justifies the small value of $\gamma$ used in our numerical explorations. 
Finally, (c) shows that the absolute value of $\phi_d$ never exceeds ${\phi_d}_\text{max}=0.75\times 3\times10^6\ \text{(V/m)}\times 4\times{10^{-3}}\ \text{(m)}\ (\epsilon_G/HU\mu)^{1/2}\approx 8.15$, owing to the dielectric-limit breakdown constraint.

\subsection{Framework for control of DNS using MPC}

Our aim is to control DNS (see Appendix~\ref{sec:appDNS}) using an MPC loop \cite{morari1988model}, as outlined in Figure \ref{fig:mpcOC}(b).
In MPC, at certain discrete timesteps, system state measurements are fed into the {\em MPC Controller}, which derives an updated set of controls, which are fed back into the system.
In industrial contexts, where only coarse-grained control is needed, the controller can sometimes use a drastically-simplified model, such as an empirical or linear model. 
In our case, the OC framework for the RDM (Section~\ref{sec:ocFramework}) is used as the MPC Controller.

Two key issues are how to select when the controls are updated, and what measurements are used from the DNS. 
Here, the controller was called at $t=0$, and at any time $t={t_R}_i$ at which the criterion
\begin{equation}
    ||h_\text{DNS}(t={t_R}_i)-h_\text{RDM}(t={t_R}_i)||_2>\ecrit
\end{equation}
was satisfied, where $h_\text{DNS}$ and $h_\text{RDM}$ are the height according to the DNS and RDM, respectively, $\ecrit$ is a constant parameter, 
and $||\cdot||_2$ is the usual $L^2$-norm. 
The control is then performed on the time interval $t\in({t_R}_i,\tend)$, with the measurements from the DNS, namely the height $h$ and the depth-integrated velocity $q$ at the grid points used by the RDM,
forming the ``initial" conditions at $t={t_R}_i$.
In experiments information is unlikely to be available at this granularity and accuracy, and the behaviour of the system with less and/or noisier information is a key future route for investigation. In particular, it is rarely feasible to observe the flux $q$, and, while obtaining effective controls by replacing $q$ by its leading order approximation ($q\approx\frac{h^3}{3}$) is possible \cite{thompson2016stabilising}, a more robust mechanism is the target of the present work.

\subsection{Results of control of DNS using MPC}

\begin{figure}[!ht]
\centering
\begin{tabular}{ccc}
\includegraphics[width=0.32\textwidth]{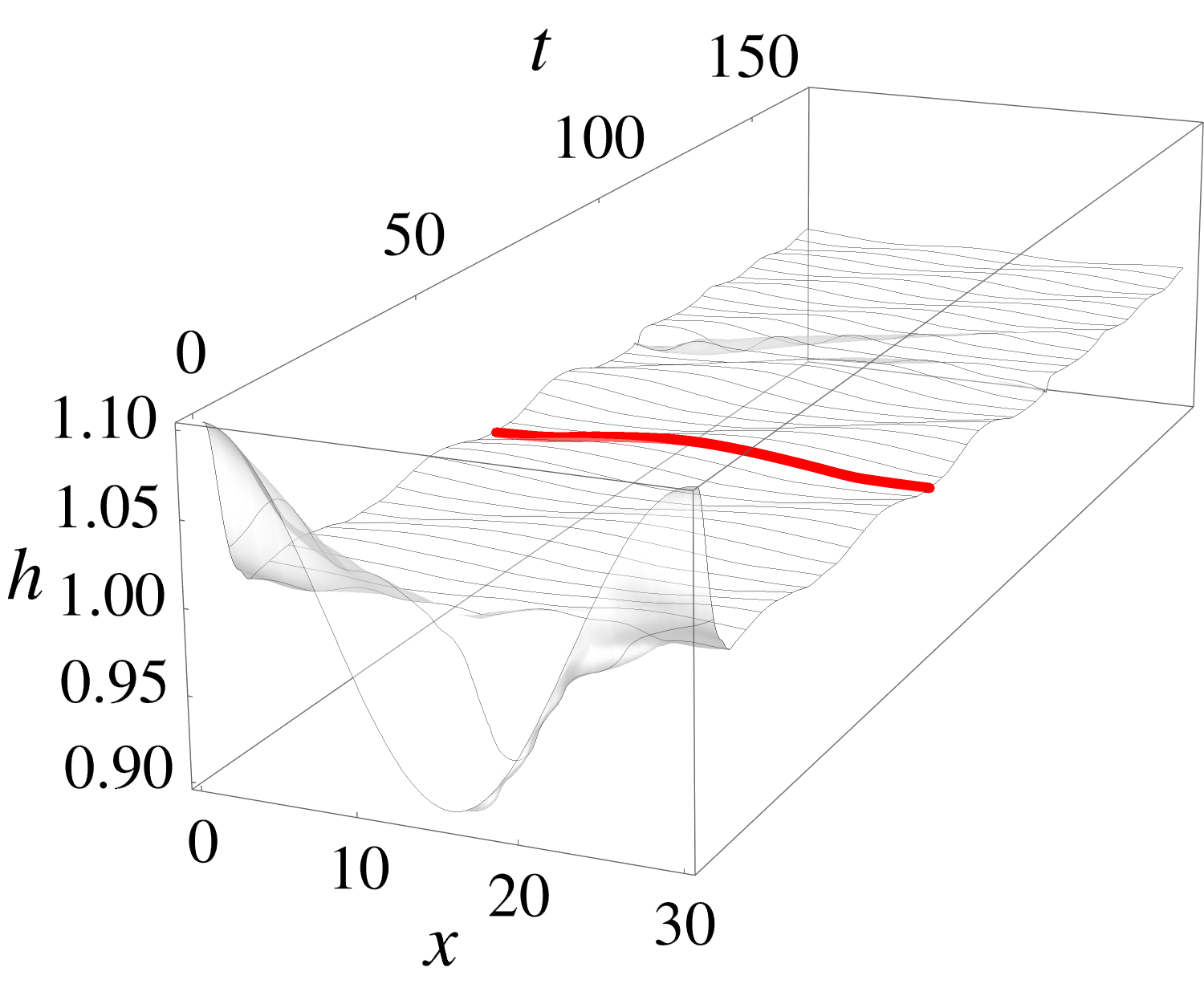} &
\includegraphics[width=0.32\textwidth]{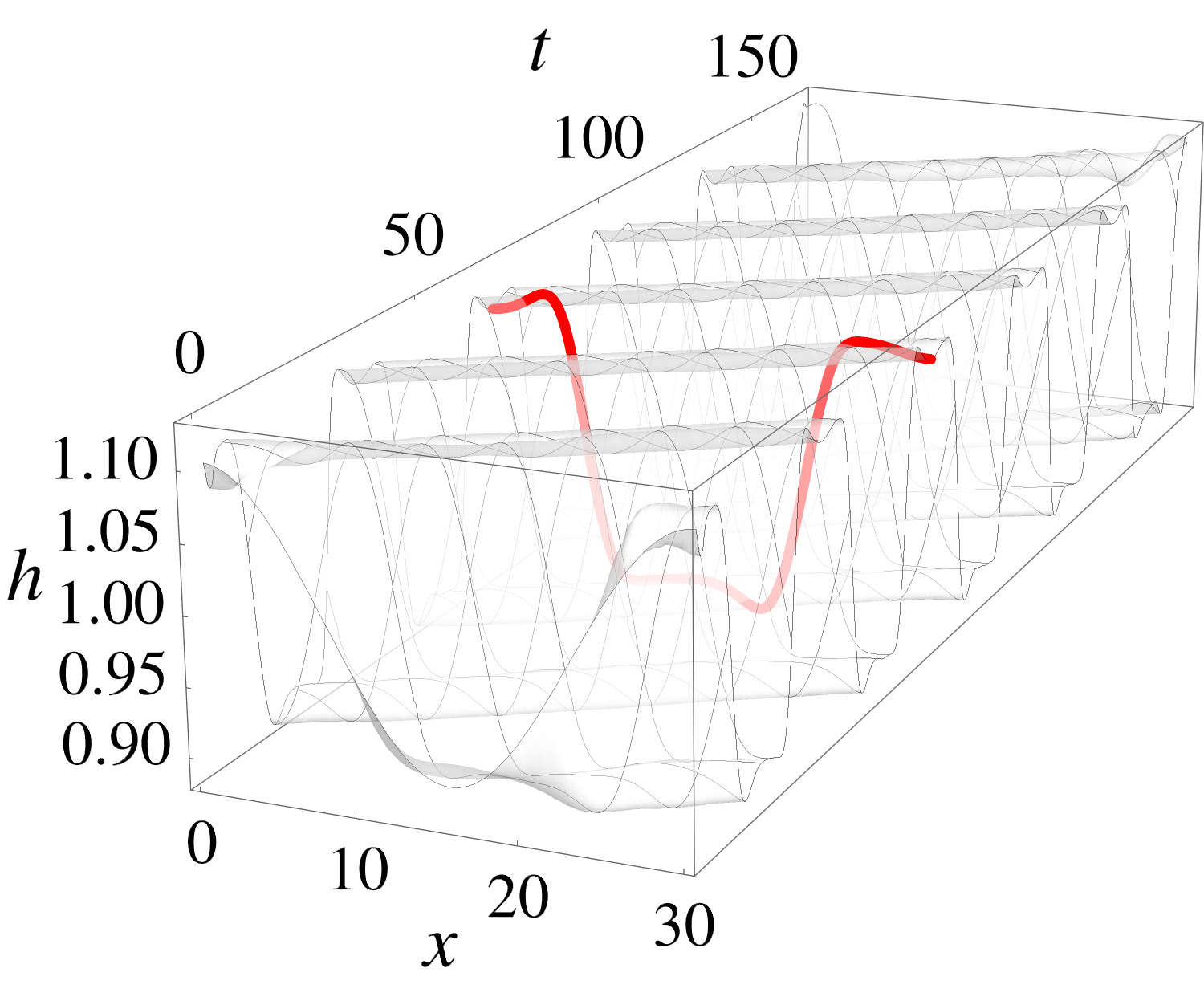} &
\includegraphics[width=0.32\textwidth]{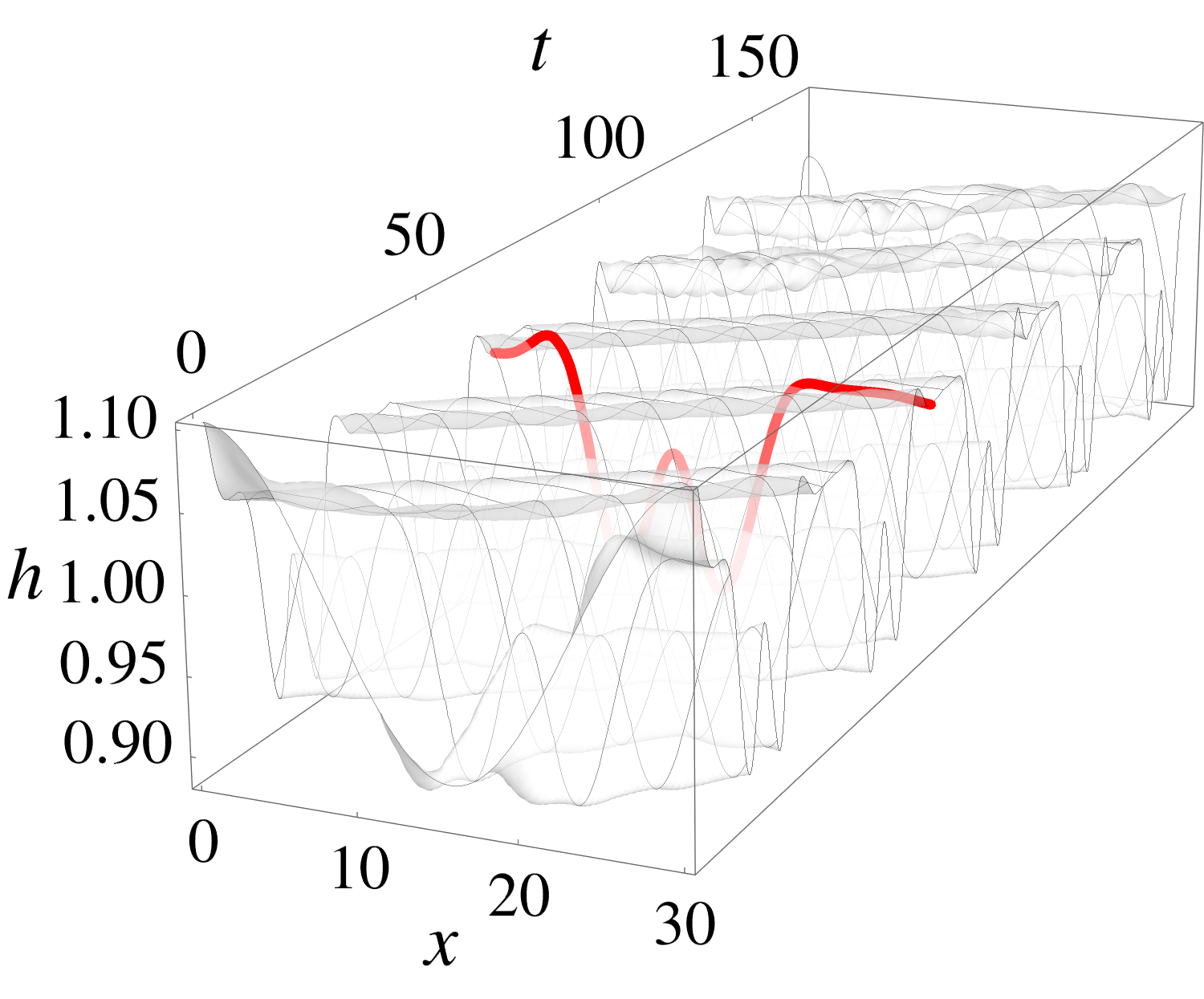} \\
(a) & (b) & (c) \\
\includegraphics[width=0.32\textwidth]{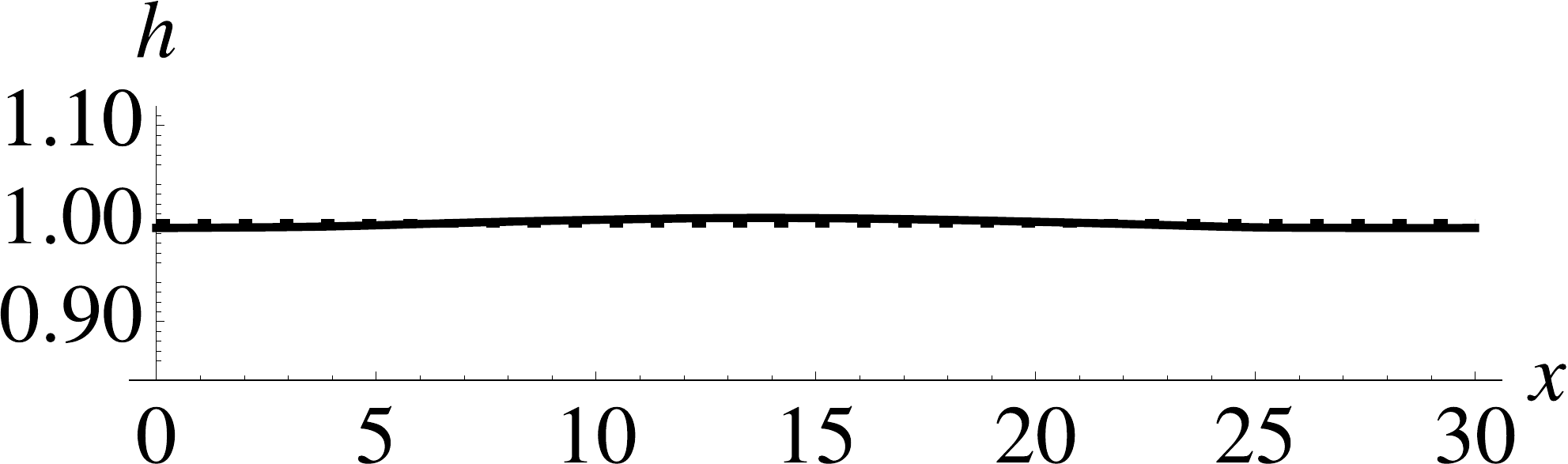} &
\includegraphics[width=0.32\textwidth]{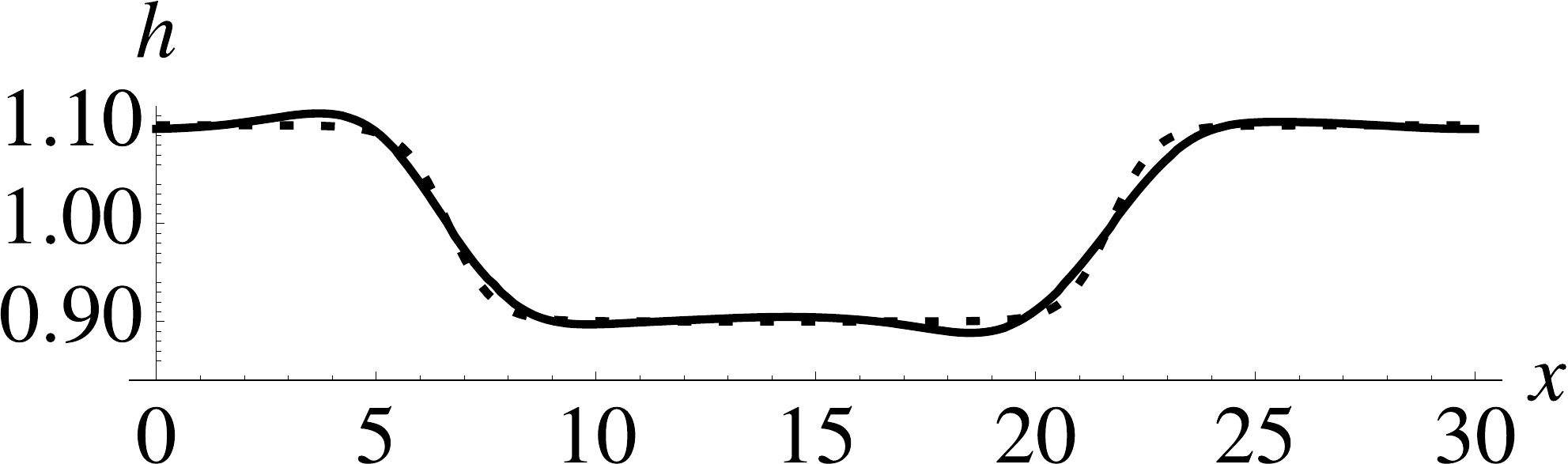} &
\includegraphics[width=0.32\textwidth]{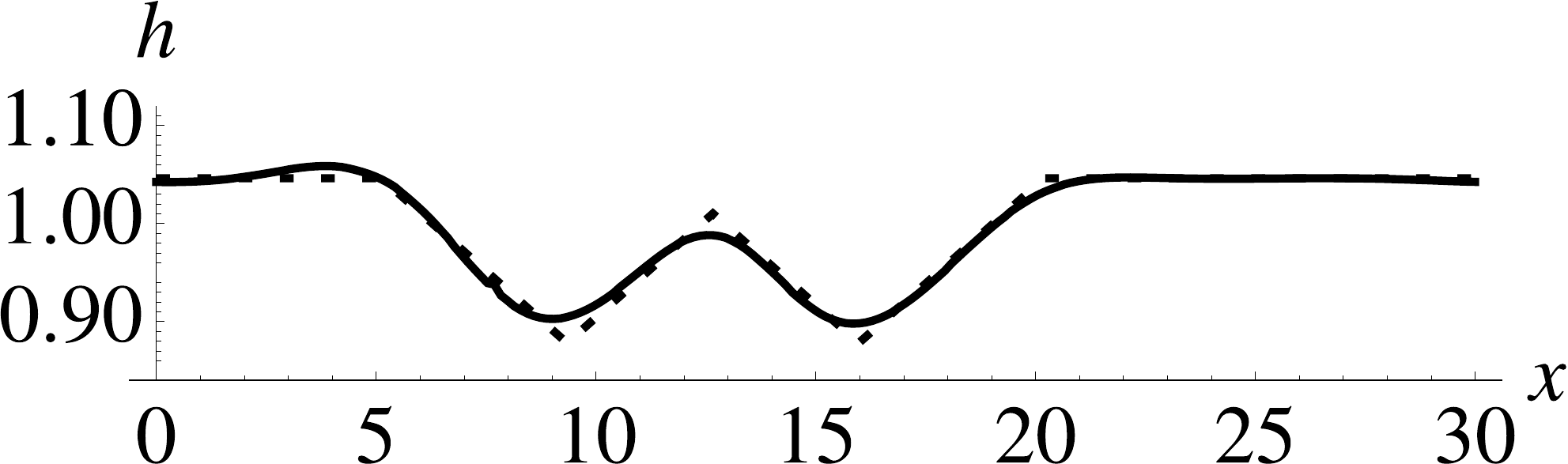} \\
(d) & (e) & (f) \\
\includegraphics[width=0.32\textwidth]{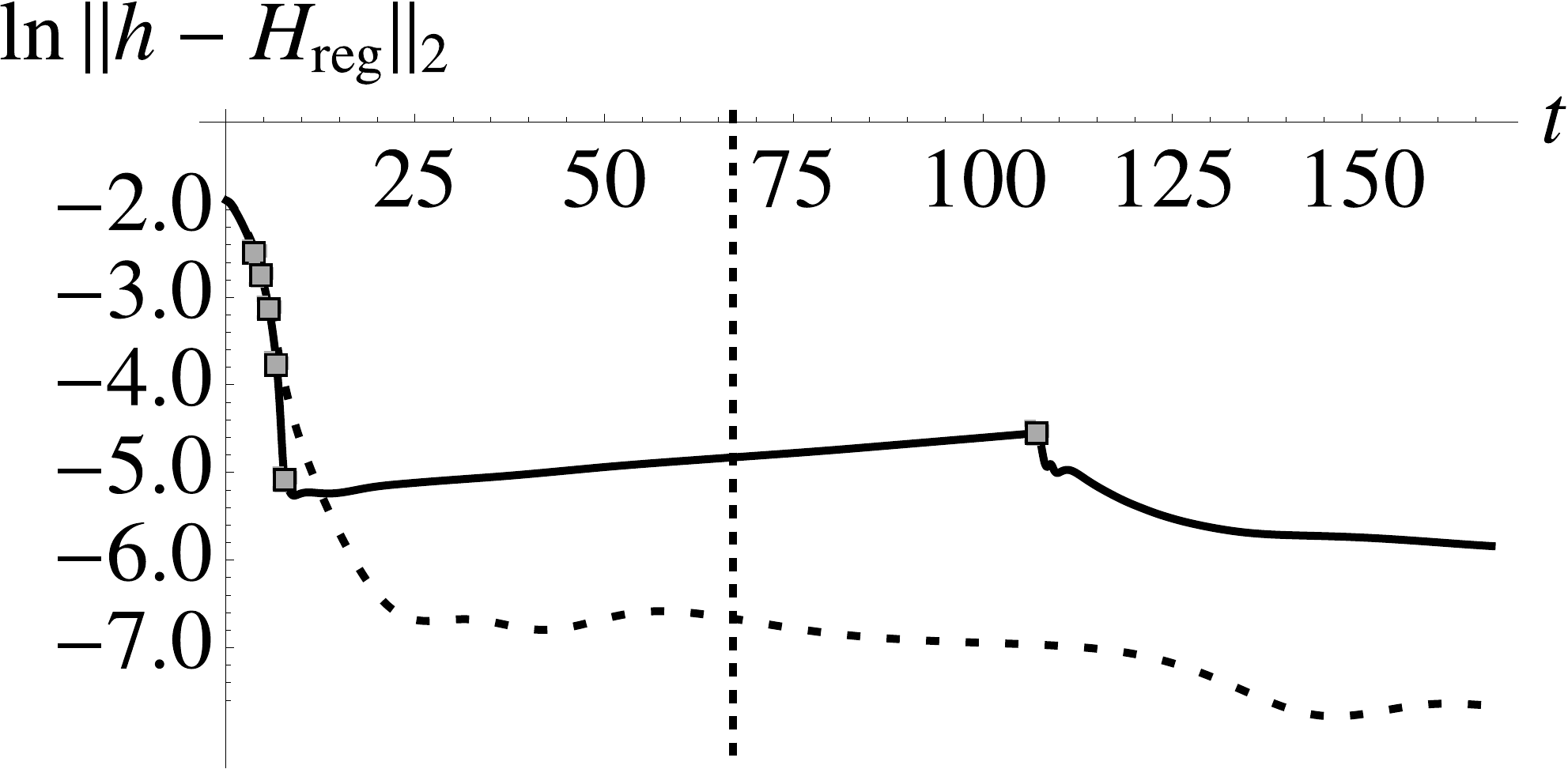} &
\includegraphics[width=0.32\textwidth]{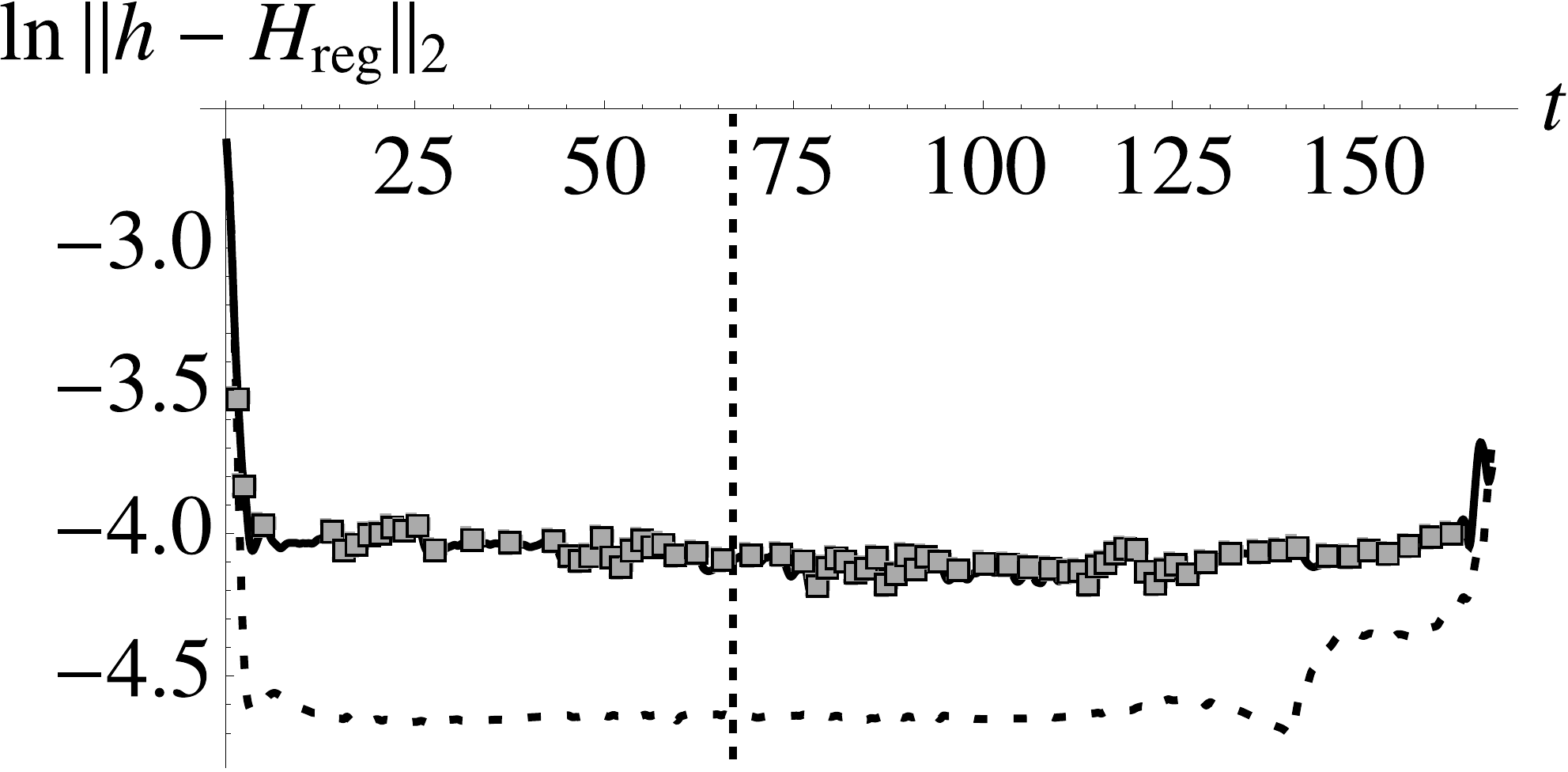} &
\includegraphics[width=0.32\textwidth]{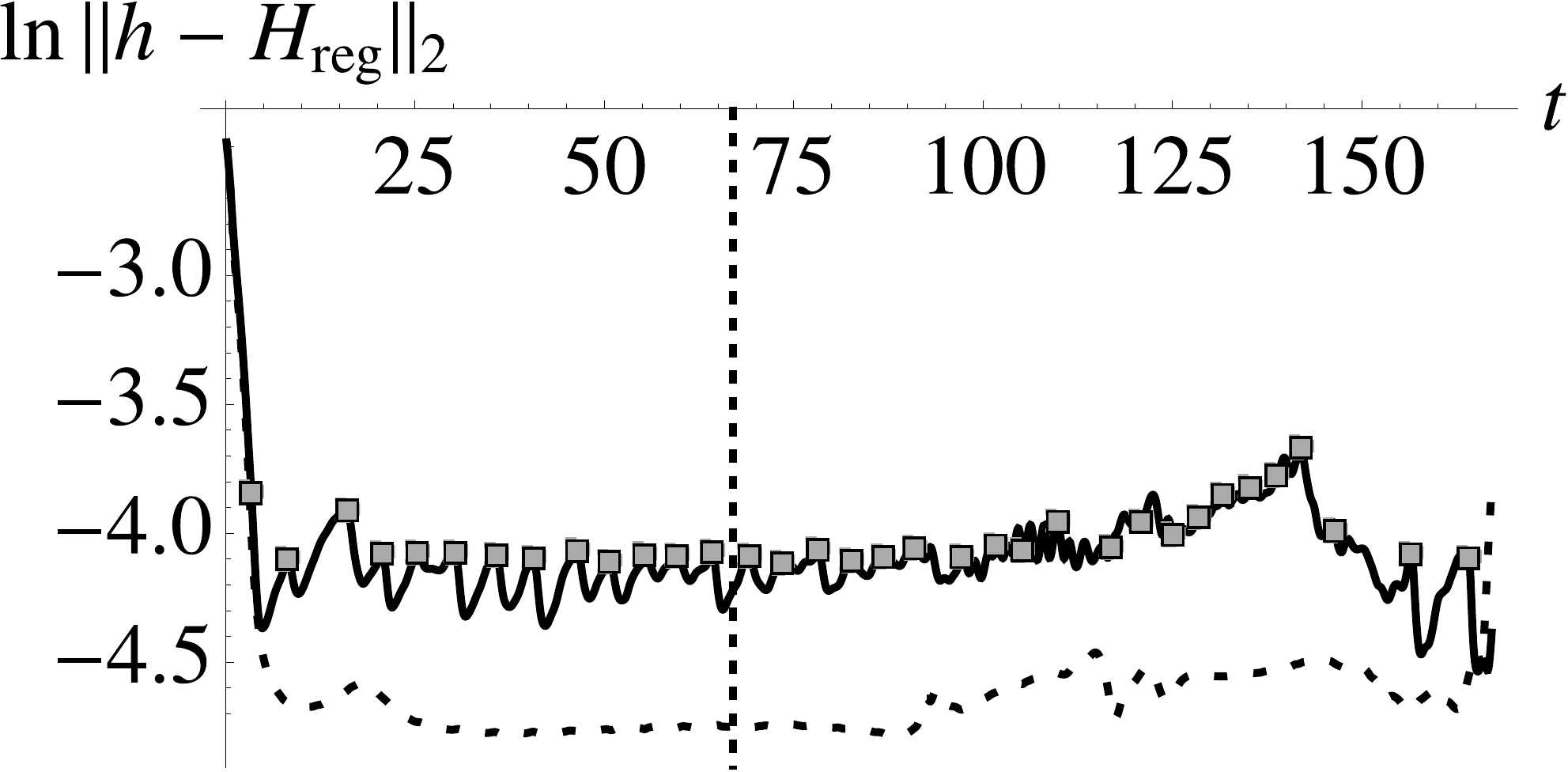} \\
(g) & (h) & (i)
\end{tabular}
\caption{Regulation MPC control of DNS towards (a), (d), (g): a uniform state ($\ecrit=
1.5\times10^{-5/2} $) ; (b), (e), (h): a smoothed top-hat shape ($\ecrit=
1.5\times10^{-2}
$); (c), (f), (i): a ``W"-like shape ($\ecrit=
1.5\times10^{-2}
$). (a), (b), (c): DNS interfaces over time. Highlighted time $t=0.4\tend$. (d), (e), (f): DNS interface (solid line) and target state (dashed) at $t=0.4\tend$. (g), (h), (i): $\ln ||H-H_\mathrm{reg}||_2$ over time; dashed line: RDM; solid line: DNS; vertical dotted line: $t=0.4t_f$. Filled symbols indicate control recomputations.}  
\label{fig:controlIms}
\end{figure} 

We examine control towards three different targets, as shown in Figure \ref{fig:controlIms}. 
Our chosen configuration is naturally unstable, with uncontrolled evolution resulting in convergence to saturated nonlinear travelling waves. 
In all three cases the DNS interface is correctly controlled towards the target shape. 
Due to the translation-agnosticism these shapes translate over the time interval - see (b) and (c). 
As expected, the match is imperfect, as both capillary and electric field effects are incompatible with corner structures.
%
Directing the flow towards a flat state necessitates few recomputations: after five initial recomputations to bring the interfacial DNS height close to flat, only one more is needed later.
For the more complex shapes the DNS and RDM diverge more rapidly, hence more recomputations are required. 

We also examined varying $\ecrit$ when controlling towards \eqref{eq:tophat}. 
When $\ecrit=\infty$, there is no MPC loop: the DNS simply uses the initially predicted control. 
%
%
A lower $\ecrit$ results in more accurate control (smaller $||h-\hreg||_2$), although this results in more recomputations. 
However, the iteration uses the existing control as its initial guess, so the small deviation (due to small $\ecrit$) typically yields rapid convergence, hence quick recomputations. 
In terms of wall clock times, computations required $2.81\times10^6$ s for $\ecrit=\infty$; $2.96\times10^6$ s for $\ecrit=1.5\times10^{-3/2}\, (2 \text{ recomputations})$; $3.19\times10^6$ s for $\ecrit=1.5\times10^{-2}\, (120 \text{ recomputations})$. 
In the cases discussed above, as well as part of broader numerical experimentation, we found that the additional control functionality incurs a computational cost which is one order of magnitude less (and most often much less) than the DNS component of the calculation, thus making it a powerful and relatively inexpensive component in the context of the full methodology. 

We thus find the proposed control mechanism to be both sufficiently robust and versatile in reaching target states using a productive interplay between modelling and computational approaches, leading to a promising framework. While deployed on a specific example, we envision generalisation to different actuation mechanisms and related multi-fluid systems to be within reach. 
Furthermore, uncertainty quantification aspects arising from issues such as interface measurement, state estimation (e.g. estimating $q$ from observations of $h$ only, or estimating $h$ from a finite number of possibly noisy observations of the interface) and actuation represent important milestones in view of specialisation towards specific applications of interest.

\begin{acknowledgments}
The authors thank Jose M. Lopez-Herrera (Univ. Sevilla) for helpful discussions on the electrohydrodynamics implementation in the DNS. RC and SNG also gratefully acknowledge the support of the EPSRC (EP/V051385/1). Data and implementations will be made available upon reasonable request.
\end{acknowledgments}

\appendix
\section{Direct Numerical Simulations}
\label{sec:appDNS}

We used the \url{http://basilisk.fr/} \cite{popinet2015quadtree} open-source package as a development platform for our proposed framework due to its accuracy, efficiency, multi-physics capabilities (electric force coupling \cite{lopez2011charge} to the fluid dynamics solver \cite{popinet2009accurate}), and ease of integration with the MPC loop. For our setup a tailored fixed grid, gradually more refined from the active electrode to the region occupied by the liquid film, was a suitably robust option given the non-local electric field effects. This resulted in $\approx 40$ cells allocated to the undisturbed film height, providing sufficient support for the nonlinear liquid film motion, with $\mathcal{O}(10^4)$ grid cells in total and $\mathcal{O}(10^2)$ CPU hour runtimes. This computational effort was typically distributed over $4-16$ CPUs, resulting in overall reasonable execution cycles. Development and validation were conducted in varied settings, and for results in Section~\ref{sec:ocFramework} the domain length was set to $L=30H$, and the domain height to $5H$, where $H$ is the undisturbed liquid film thickness. 

\section{Numerical solution of the optimal control problem}
\label{sec:appOC}
The equations for the RDM \eqref{eq:finalKinEq}, \eqref{eq:finalEvEq} and \eqref{eq:fInnerCons}, and corresponding adjoint equations,
\begin{align}
\nonumber 0 = &-\lambda^h_t+2\gamma_\text{reg}\left(H-h\right)-Re\left(\frac{9}{7}\frac{q^2}{h^2}\lambda^q_x+\frac{1}{7}\frac{\lambda^q q}{h^2}q_x \right)+\frac{5}{6}\sin\alpha\,\lambda^q +\frac{5}{6}\cos\alpha \,h\lambda^q_x\\
&\quad -\Gamma \left[\frac{5}{2}\lambda^q_x h_{xx}+\frac{5}{2}h_x\lambda^q_{xx}+\frac{5}{6}h\lambda^q_{xxx} \right] -f\lambda^f+\frac{f}{2}\left(d-h\right)^2\lambda^f_{xx}+E_b\frac{5}{6}f\lambda^q f_x\\ & 
\quad + 4\frac{ (q+\lambda ^q) }{h^2}h_x \lambda^q_x
-4\frac{q \lambda ^q}{h^3}h_x^2
+4\frac{q \lambda ^q}{h^2}h_{xx}
-\frac{15}{2h}q_x\lambda^q_x -6\frac{ q }{h}\lambda^q_{xx}
  -\frac{3}{2}\frac{\lambda ^q }{ h}q_{xx}
+5\frac{q \lambda^q}{h^3},\nonumber\\
0=&-\lambda^q_t+\frac{5}{2Re}\frac{\lambda^q}{h^2}-\frac{1}{7}\frac{\lambda^q q}{h^2}h_x+\frac{1}{2Re}\frac{\lambda^q}{h^2}h_x^2+\frac{1}{Re}\lambda^h_x-\frac{17}{7}\frac{q}{h}\lambda^q_x-\frac{9}{2Re}\frac{1}{h}h_x\lambda^q_x+\frac{3}{2Re}\frac{\lambda^q}{h}h_{xx}-\frac{9}{2Re}\lambda^q_{xx},\\
0=&\frac{5}{6}E_b\left(\lambda^q h\right)_x+\lambda^f\left(h-d\right)+\frac{1}{6}\left(d-h\right)^3\lambda^f_{xx},
\end{align}
determined from \eqref{eq:adjoinEqs}, were solved using a fully-implicit centred-finite-difference C++ code that has been extensively tested in other film problems \cite{wray2017accurate,wray2017reduced}. We used $200$ points in space and a fixed time step to facilitate the OC code, resulting in $1000$ in time. The code was optimised to allow a forward simulation to run in under one second on a single core of an Intel Core i5-10400.

\bibliography{ocbiblio}

\end{document}